%Paper: hep-th/9308037
%From: PTVP@IBM.RZ.TU-CLAUSTHAL.DE
%Date: Sun, 08 Aug 93 23:19:42 MET

\magnification 1100

  \def\zc{\gamma}

 \def\zD{\Delta}

  \def\hh{\hat X^0}
\def\p{\partial}

% caligraphic:

 \def\cA{{\cal R}}

     \def\cO{{\cal O}}  \def\cP{{\cal P}}
       
 \def\cU{{\cal U}}

% doubled letters:

     \def\dZ{Z\!\!\!Z}

  \def\hf{{\scriptscriptstyle (-)}}
 \def\sint{{\textstyle\int}}

\def\jj{X}  \def\j0{X_0}

\def\tsug{T^{(\rm sug)}} \def\ttot{T^{(\rm tot)}} \def\tred{T^{(\rm f\, f)}}
 \def\ltot{L^{(\rm tot)}} \def\lred{L^{(\rm f\, f)}}

\def\ii{{\bf I}} 
\def\lbt{L} \def\tbt{T}
\def\kk{{t}}
\def\hs{\widehat {sl}(2)}

\def\fz{f_1}
\def\fy{f_0}

\nopagenumbers

\centerline{ }
\vskip 3cm
\centerline{\bf  VIRASORO SINGULAR VECTORS VIA QUANTUM DS  REDUCTION
\footnote{${}^*$}{\rm Work  supported in  part by the Bulgarian Foundation for
Fundamental Research
 under contract $\Phi -11 - 91. $}}
\vskip 2cm
\centerline{\bf A.Ch. Ganchev  and V.B. Petkova}
\bigskip
\centerline{ Istituto Nazionale di Fisica Nucleare, Sezione di Trieste,
Italy, }
\smallskip
\centerline{  Institute for Nuclear Research and Nuclear Energy,
Sofia 1784, Bulgaria \footnote{$^{**}$}{\rm Permanent address.}}
\vskip 2cm

\centerline{\bf Abstract}
\vskip 1cm

The  BRST quantisation of the Drinfeld - Sokolov (DS)
reduction is exploited to
recover all singular vectors of the Virasoro algebra Verma modules from the
corresponding $A^{(1)}_1\,$ ones.
The two types of singular vectors are shown to be identical modulo
 terms trivial in the $Q_{{\rm BRST}}$ cohomology.
 The main tool is a quantum version of the DS gauge transformation.
\vskip 1cm

\vfill May 1993
\vfill\eject

\pageno=1\footline{\hfil\tenrm\folio\hfil}

\noindent{\bf 1.}
 In [1] Bauer et al (BFIZ) interpreted any of the Virasoro singular vectors
found in  [2]
  as a  scalar Lax operator [3]  recovered by a matrix valued system,
associated to a finite dimensional representation of $sl(2)$.
In [4], [5]  it  was shown that
the BFIZ matrix system emerges naturally
 from a system generated by the $\hs$ Knizhnik -- Zamolodchikov
(KZ) equation. Namely the two are related by a  quantum DS -  type gauge
transformation.  Furthermore   a
conjecture was made [5] relating the general  singular vectors of the Vir
 Verma modules
to the  Malikov -- Feigin -- Fuks  (MFF) [6]
$\hs$  singular vectors.
 An explicit algorithm based on it was shown to reproduce
the simplest examples. The technique in [4], [5]
was based on a   realisation of the generators
by  differential operators and on  a
proper basis suggested by general explicit solutions of the KZ equation.

Here we reformulate the problem   in purely algebraic terms,
implementing the
quantum DS reduction in a BRST framework [7] --  however for
Verma, instead of Fock  modules (an approach advocated also in
 [8]). This  enables us in particular to
 prove the conjecture in [5]  exploiting the explicit formula
 for the Vir singular vectors written by Kent [9], thus demonstrating that
the formulae in [6] and [9]
   are not just analogous in form,  but up to BRST trivial terms,
simply identical. In a forthcoming paper [10] the approach developed here
will be generalised for the $W_3$ - algebras.

\medskip\noindent{\bf 2.}
For a field $A(z)\,$ one has the mode expansion
$\,  A(z) = \sum_{m\in\dZ} A_m \, z^{-m-\zD_A}\,.$
 The (anti)commu\-ta\-tion relations of the modes are equivalent to the
singular part of the operator product expansions (OPE) of the fields.
The normal ordered product $(A\,B)(w)$ of two fields $A$ and $B$ is
by definition the zero order term in the operator product
expansion (OPE) of $A(z)\,B(w)$. We will
 refer to  [11]  for rules to work with OPEs.
With dot or $\p $ we denote a derivative; $(\dot\psi)_m = -(m+\zD) \,\psi_m$.

The affine Lie algebra $g = \hat{sl}(2)_k$
is defined by the OPE
$$\eqalign{
  \jj^+(z) \, \jj^-(0) &= {k \over z^2} + {2\jj^0(0)\over z} + \dots \,,
\cr
  \jj^0(z)\, \jj^{\pm}(0) &=  {\pm \jj^{\pm}(0)\over z}  + \dots \,,
\cr
  \jj^0(z)\, \jj^0(0) &= {k/2 \over z^2}  + \dots \,,
\cr}\eqno(1)$$
assuming $ \zD_X = 1$.
Next we need a system of fermionic ghosts, i.e.,
anticommuting fields $b,c$ with $\, \zD_c = 1 \,, \, \zD_b =0 \,,$
$$
  b(z) \, c(0) = {1\over z} + \dots \ \ \,. \eqno(2)
$$

\medskip

 The constraint
 $X^+(z)=1$ is introduced in a BRST fashion. The BRST charge $Q$ is [7]
$$  Q = \oint_{C_0} dz\,\,(cX^+ -c\,)(z) = (c\,X^+)_{-1} - c_0 \,,  \quad
Q^2=0\, .\eqno(3)$$

The Sugawara energy momentum tensor $ \tsug=
 ( X^a X_a) \,/ 2(k+2) $ is extended to a field $\ttot$
commuting with $Q$,

$$\ttot=
 \tsug (z) +   \dot X^0 (z)  +  (\dot b \, c) (z) \, .\eqno(4)$$

Let us introduce the field [12]
$$ \hh = X^0 + (b\,c) \,.\eqno(5)$$
Its operator product with $X^{\pm}$ is the same as that of $X^0$,
while
$$  \hh(z) \, \hh(0) = {1/ 2\nu  \over z^2} +\dots \, \, ;\qquad \, 1/
\nu = k+2 \,. \eqno(6)$$
The energy - momentum tensor

$$\tred =  \nu  (\hh \hh) +(1- \nu) \p \hh
\,, \eqno(7)$$
has conformal anomaly $c_{\nu }= 13- 6 \nu  - 6/ \nu$.

By a  straightforward computation using
the technique of [11] one can show that
the  tensor $\ttot$ can be rewritten in terms of $\hh$ as

$$
  \ttot  =   \tbt +  \nu \{ Q, (b X^-) \} \, , \eqno(8)$$

$$\tbt =  \tred + \nu X^- \,. \eqno(9)$$
Any of  the three tensors $\ttot$, $\tred $, and  $ \tbt $
 closes a Virasoro algebra with one and the same
value of the  central charge $c_{\nu }$. Furthermore
$$ \left( \ttot (z)\,  - \, \tred (z) \right)
\, \hh  (0) = \, ... \  ({\rm regular})
 \,.\eqno(10)$$

 The tensor $\tbt\, $ appears also in the
recent paper [13], where it is argued
  that the universal enveloping algebra of
the  Vir algebra of $\tbt$
appears as the (only nonzero)
cohomology of the complex generated by the action of $Q$  in
(the completion of) $\, \cU_g \otimes {\rm Cliff}(b,c)\,.$
\bigskip
\noindent{\bf 3.} Now let us turn to representations.
Let $ V_0 = V_{\{j,\nu\}} $
be an $\widehat{sl}(2)_k$ Verma module highest weight vector
of weight $r$ (spin $j=r/2$), i.e.,
$$
2  X^0_0 \, V_0  =  r \,  V_0  \,,\qquad {\rm and}\qquad
  X^+_m \, V_0  = X^-_n \, V_0  = 0 \,\ {\rm for}\,\ m \ge 0, n > 0 \,.
\eqno(11)$$

Denote $ V_t = (X^-_0)^t\, V_0  \,.$
Obviously $2 X^0_0 \, V_t = (r-2t)\,V_t$
and $\j0^+ \,V_t  =\zc_t(r) \, V_{t-1}\,, \, \zc_t(r) = t(r+1-t)
\, $. Thus if $r$ is a non-negative
integer
$ V_{r+1} $ is a singular vector in the Verma module. For a while
we restrict ourselves with this simplest
 subseries of $\widehat sl(2)_k$ singular vectors.

\medskip
Assume that $ V_0 $ is also a vacuum vector for the $b,c$ system, i.e.
$$
  b_m \, V_0  = c_n \, V_0  = 0 \,\ {\rm for}\,\ m>0, n\ge 0 \,.
\eqno(12)$$
Obviously $ Q\,V_t = c_{-1} \j0^+\, V_t$ and therefore $Q$ annihilates
exactly the singular vectors $V_{r+1}$ of the $\hs$ Verma module.
Furthermore from the explicit expressions (7,8,9) for  $\tbt$ and $\ttot$
$$\ltot_n\, V_t  =\lbt_n\, V_t
 = \lred_n\, V_t = 0  \,,\ \  {\rm for} \ n\ge 1\,, $$
$$  \ltot_0 \, V_t  = \lbt_0 \, V_t
 =\lred_0 \, V_t =  h_{\{j-t;\nu \}}\, V_t
 \,, \quad h_{\{J;\nu\}}=\nu J(J+1) -J =
 h_{\{-J-1+1/\nu ;\nu\}} \,. \eqno(13)$$

The basic relation (8) can be solved for $X^-$ and furthermore
recursively for any power of $X^-_0$. Namely first splitting

$$
  \{Q,(bX^-)_n \} \,V_p = \left( \{Q,\sum_{m=0}^n b_{-m}\,X^-_{m-n}\}
   + X^-_{-n-1} X^+_0 \right) \, V_p
$$
we rewrite (8) applied to   $V_p$ as

$$
  X^-_{-n}\,V_p = \left[
 {1\over \nu}\,\left(\ltot_{-n-1}-\lred_{-n-1}\right) -
  \{ Q, \sum_{m=0}^n  b_{-m}\,X^-_{m-n}\} \right] \, V_p
  -X^-_{-n-1} \,X^+_0 \, V_p \,.
\eqno(14)$$
This  is the analog of
 the triangular KZ equation exploited in [5].
Thus, analogously to what has been done there, one
 can eliminate recursivelly all $X^-_{-n}$ with $n>0$ obtaining
$$
 X^-_0 V_p = \sum_{n=0} \left[
\,  {1\over \nu}\left(\ltot_{-n-1}-\lred_{-n-1}\right) -
  \{ Q, \sum_{m=0}^n  b_{-m}\,X^-_{m-n}\} \right] \, (- X^+_0)^n
   \, V_p  \,.
\eqno(15)$$

For $p=r$ the relation (15) expresses the $\hs$ singular vector
$V_{r+1}$ in terms of the Vir generators $\ttot$. To get rid of
the remaining dependance on Heisenberg algebra generators present
in $\lred$ we shall exploit as in [5] a ``gauge'' transformation --
a quantum version of the classical DS gauge transformation. To do
that  let us first  introduce an auxiliary $sl(2)$ algebra
 with generators $\kk^{\pm,0}$
having the same commutation relations as the zero modes $X^{\pm,0}_0$.
Let $U_0$ be a highest weight vector of a finite dimensional, spin $r/2$,
auxiliary $sl(2)$ module, thus $\kk^+ U_0 = 0 =  (\kk^-)^{r+1} U_0 $.
Consider
$$
  \ii = \sum_{n=0}^r \,(X^-_0)^n \, V_0 \otimes(\kk^-)^{r-n}\, U_0 \,.
\eqno(16)$$
Obviously (skipping for short direct product notation)
$$
  (X_0^+ -\kk^+) \,\ii =0,\qquad
  (X_0^- -\kk^-) \,\ii = (X_0^-)^{r+1}\, V_0 \otimes U_0  ,\qquad
  (X^0_0 +\kk^0) \,\ii =0 \,.
\eqno (17)$$

Tensoring (15) by  $\ \otimes(\kk^-)^{r-p}\,U_0$, then  summing
 over $p=0,\dots,r$ and using  (17) we obtain

 $$ (X_0^-)^{r+1}\, V_0 \otimes U_0  =
  \left( -t^- + {1\over \nu} \sum_{n=0}  (- t^+)^n
 \left(\ltot_{-n-1}-\lred_{-n-1}\right) \right) \, \ii -
   \{ Q, B \}  \,  \ii  \,,
\eqno(18)$$
where for short
$$
  B = \sum_{n=0} \sum_{m=0}^n b_{-m} \, \jj^-_{m-n} \, (-\kk^+)^n \,.
\eqno(19)$$
\medskip
\noindent{\bf 4.} Consider
the operator
$$
  \cA(u) =\, {\scriptstyle {\circ \atop \circ}}
 \, \exp\big(\sint \hh_\hf (-u) \, du \big)\,
{\scriptstyle {\circ \atop \circ}}
  \,  \equiv \, \sum_{n=0}^\infty \cA_{-n} \, (u )^n \,,\eqno(20)
$$
where $  \cA_{-n}$ are determined recursively according to
$$   k  \cA_{-k}=\sum_{l=0}^{k-1}(-1)^{k+l-1}
  \cA_{-l}\hh_{-k+l}  \, ,\ \quad   \cA_0=1\,, \eqno(21)$$
and the parameter $u$ will  be identified with   $X^+_0$ or $t^+$
in what follows.
 By  $X^0_\hf$ in (20) is denoted the holomorphic part of the field
$X^0(u) = \sum_{n\in\dZ} X^0_{-n} u^{n-1}$, i.e. $(X^0_\hf)_n=X^0_n$
if $n\le -1$ and zero otherwise.
Then for $u$ commuting with $\hh$,  e.g., $u= t^+\,$,   (20) is a true
exponent and the dots can be omitted.

{}From
$$[\hh_{-n} , Q] =- [\hh_{-n} , c_0] =  c_{-n}$$
it follows inductively that
$$
  [ Q , \cA(u) ] = - \cA(u) \, u \, c_{-1} \,,
\eqno(22)$$
and hence
$$Q \, \cA ( X^+_0) \, V_p=0 \,,  \eqno(23)$$
showing that $ \cA(X^+_0) \, \ii = \cA(t^+) \, \ii= \cA \, \ii$
 is in the kernel of $Q$.

\medskip
Next we prove the analog of the result in [5]:
\medskip\noindent{\bf
The Virasoro singular vectors of Benoit Saint-Aubin (BS-A)
for $\ttot$ are equivalent, up to $Q$ exact terms,
to the $\hs$ singular vectors $V_{r+1}.$}
\medskip
Since the BS-A vector (see (29) below)
 is recovered by the BFIZ system it is enough
to show that
 \footnote{$^1$}{ In  [5] the analog of (24) was written in terms
of the improved tensor instead of  $\ttot$. The price of avoiding
the ghosts was that only  ``half'' (the
nonnegative modes) of the Vir algebra was
recovered.}
$$
  V_{r+1}\otimes U_0 =
   \left(-\kk^- + {1\over \nu }\, \sum_{n=0} \ltot_{-n-1} \, (-\kk^+)^n \right)
  \,\cA\, \ii    + Q\,\cdots \,. \eqno(24)$$
\noindent{\bf Proof: \ \  }
Apply on both sides of
 the  equation (18)
 the gauge transformation  $\cA =
\cA( t^+)$
 from the left. The l.h.s. remains unchanged.
According to (10)  the gauge transformation
 commutes with the second term in (18). The result for the
 first term is collected in  the following
relation

$$  [ \cA(t^+) , - \kk^- ]\, \ii = {1\over \nu }
  \sum_{n=0} (-t^+)^n \lred_{-n-1} \cA \, \ii\,.\eqno(25)$$
Thus the corresponding term in (18) is cancelled leaving only the
piece
with   the generators $\ltot_{-n-1}$.

 The proof of (25) consists of several steps: 1) perform the commutator
 of the auxiliary $sl(2)\, $ generators, 2) use twice the defining
 recursion relation (21) to create a quadratic in $\hh\, $
term, 3) trade the auxiliary generator $\,t^0\,$ for
$\,X^0_0\,$ exploiting the last equation in (17),  4) group
the various terms to recover the normal product in (7), taking also into
 account (6).

Finally one checks that $\cA$  maps the last term in (18) into
  the image of $Q$.
Indeed  from the definitions it follows that
$Q\,\ii = c_{-1}\,X^+_0 \,\ii =  c_{-1}\,\kk^+ \,\ii$,  and taking
into account (22), (19)  we have
$$
  \cA \, \{ Q, B\} \, \ii =
    [ \cA , Q]\, B\, \ii +  \cA \, B \,Q \,\ii +  Q \, \cA \, B \, \ii$$
        $$= \cA \, \{ c_{-1} , B \} \, \kk^+ \, \ii+  Q \, \cA \, B \, \ii
\, =\,   Q \, \cA \, B \, \ii  \,.  \eqno(26)$$
Neglecting the terms in the image of $Q$  turns (24) into the
system of [1] with basis vectors which can be recovered from
 $\cA \, \ii \, $ using (16,17).
\medskip

\medskip\noindent{\bf 5.} There is yet another way
of recovering  the BS-A vectors from
the  $\hs$ singular vector $V_{r+1}$.
Obviously  (24)
remains true if the generators $\ltot$ are replaced by $\lbt$
 from (9) since $\cA\, \ii$ is annihilated by $Q$.
 Recall that $\tbt$
also commutes with $Q$ and provides the same Vir algebra as $\ttot$.
More than that,  this alternative expression can be derived directly
in terms of the simpler tensor $\tbt$ - avoiding the iteration of
(8) described in (14-15, 18).

Indeed identify the parameter $u$ with $X^+_0$ ,
 instead of $t^+$, i.e., from now on
 $\cA \equiv \cA(u) = \cA( X^+_0)$.
Then one has for any vector  $V$ annihilated by all positive grade
generators as well as by the ghost zero mode $c_0\,$

$$\cA\, X^-_0\, V =
 {1\over \nu } \sum_{p=0}
\,\lbt_{-p-1}  \, \cA \,\, (-X^+_0)^p\, V
 \,.\eqno(27)$$
The proof of (27) is similar to that of (24). The
  $\lred\,$ - piece of $\lbt\,$  (cf. (9))
 is recovered by the commutator of $X^-_0$ with $ (X^+_0)^k$ . Its derivation
repeats that  of (25),  replacing everywhere
$t^{\pm}$ with $X^{\pm}_0$.  The change of sign is compensated by
the reshufling of the generators. The
negative modes of $X^-\,$
 come from the commutator of $X^-_0$ with $\cA_{-k}$ .
 One has to  exploit the following  relation proved by induction
$$ [\cA_{-k}, X^-_{-n}] = \sum_{l=0}^{k-1}(-1)^{l+k}
X^-_{-k-n+l} \, \cA_{-l} \,.\eqno(28)$$
The relation (27) (and the related to it expression
  written in terms of the auxiliary $sl(2)\,$ algebra )
 can be
interpreted as the quantum version of the
DS gauge fixing transformation of the
constrained system, leading to the classical (gauge invariant)
 counterpart of the current $\tbt(z)\,. $

Now choose  $V=V_t$. Then the power of $X^+_0$ in the
r.h.s. of (27) produces
 $\, X^-_0 V_{t-p-1}\, $ times the  constant
 $\,(-1)^p \zc_t(r) ....\zc_{t-p+1}(r) $
 and we can move  again $\cA\, $ to the right using (27) - repeating
 this until $V_0=\cA\, V_0 $ is reached. This reproduces the vectors $F_{t+1}
\equiv \cA\, V_{t+1}\, $
of [1]  expressed as
functions of the generators $L$, and for $t=r$ we  immediately ``generate''
 in this way directly the BS-A singular vector, i.e.,

$$V_{r+1} = \cA \, (X^-_0)^{r+1}\,V_0 = \cO_{(r+1;\nu)}   V_0$$
$$=
 \prod_{i=1}^{r} \zc_i(r) \sum_{s=1}^{r+1} (-{1\over \nu})^s
  \sum_{k_i\geq 0: \sum_{i=1}^s k_i = r+1-s}
  { L_{-1-k_s} \dots L_{-1-k_1} \over
  \zc_{k_1+\dots+k_{s-1}+s-1}(r) \dots \zc_{k_1+k_2+2}(r)
  \zc_{k_1+1}(r) } \,V_0\,.
\eqno(29)$$
 e.g.,  for $\,  t=1\,,$
$$\cA \, V_2 =  {1\over \nu }\lbt_{-1}\,  \cA\, V_1 -  {1\over \nu }
\lbt_{-2}\, X^+_0\, V_1 = ( {1\over \nu })^2 \left(( \lbt_{-1})^2
-\nu \zc_1(r)  \lbt_{-2} \right) = F_2  \,,$$
which recovers the Vir singular vector  if $r=1$.
 More generally this converts  any element in the
kernel of $Q$  of the type $ \cA\, V_t \,, \, t=1,2,..., r, r+1, ...
 \, $ --  into an element of the Vir module  generated on $V_0$.

Since  $F_0= \cA\, V_0 =
 V_0\,, \, \,  F_{r+1} =\cA\, V_{r+1}= V_{r+1} \,,$
the proof that (29) is a Vir singular vector is an immediate
consequence of  (13), i.e., we have
 $$\lbt_n\,  \cO_{(r+1;\nu)} V_0 =0=\lbt_n \, F_0\,, {\rm for}\ \  n\ge 1
 \,, \ \
\lbt_0\, F_0 = h_{\{j;\nu\}}\,F_0\, ,
 \  \ \lbt_0\,  \cO_{(r+1;\nu)} V_0 = h_{\{-j-1;\nu\}}\,
 \cO_{(r+1;\nu)} V_0\,.  \eqno(30) $$
Given (27) and  the auxiliary $sl(2)\,$
algebra (16,17),  one can  get back the BFIZ system written as in
(24), but with $\ltot\,$ replaced by $\lbt\,$ and
  the $Q$ - exact term
in the r.h.s. of (24) dropped.

\medskip
Obviously the second method of deriving the
singular vectors of the Virasoro algebra
 is technically simpler since it does not require
the knowledge of the more complicated $\ttot$ (nor its recasted
form (8)). This is important
for  generalisations  to other $W_N$ algebras -- especially in
view of the recently proposed
 general algorithm of [13] for constructing the analogs of $\tbt$.
 This will be  demonstrated for the  $W_3 $ -  algebras in [10].

\medskip

\medskip\noindent{\bf 6.}
Up to now we have considered only a subclass of the $\hs$ singular
vectors.
 Let us now give an idea
how one can  recover the general Vir singular vectors
in the form proposed by Kent [9].

The explicit expression for $ \cO_{(r+1;\nu)} \,$  can be
reordered [9], pushing all generators $\, \lbt_{-1} \,$ to the left,
 as a sum of terms of decreasing powers of
$\, \lbt_{-1} \, $, i.e., terms of the type
$$
{1\over  (\nu )^{r+1 } }\, c_{i_1,i_2,..., i_s} (\nu, r) \,
 \lbt_{-1}^{r+1-l} \lbt_{-i_1} ...\lbt_{-i_s}
\,, \ \qquad
\sum_{t=1}^s  i_t =l \,, i_t \ge 2 \,, l= 0, 2,3,  ... , r+1\,.$$
The coefficients
$\,c_{i_1,i_2,..., i_s} (\nu, r) \,$ depend polynomially on the parameters
$\, \{ \nu\,, r \} \,,$ and so they admit an
 analytical continuation to arbitrary values of $r$, when the series
representing  $ \cO_{(r+1;\nu)} \,  $   becomes
infinite.
Then $ \cO_{(r+1;\nu)} \, V_0 $
is formally a singular vector (i.e, it satisfies the conditions (30)),
 or more precisely it is a singular vector  in the (generalised)
 Verma module of the algebra obtained by extending Vir with the powers
 $(\lbt_{-1})^a$ with arbitrary value of $a$. It was shown in [9] that
proper compositions of such operators provide  true singular vectors
in the Vir Verma modules.

The construction in [9] was inspired by the analogous
expressions for the general  singular vectors in the Verma modules of
$\, A^{(1)}_1\,$ in [6].
Let $V_0=V_{\{ J\,; \nu \}}\,,$ for  $k\not =2\,,$
 be a  $\hs_k$ Verma module h.w. state
 of spin $J\,$ and assume that it furthermore
obeys the conditions (12).  Suppose
 that the weight $\ 2J+1$ can be written (not uniquely in general)
as  $\ 2J+1 = m-(m'-1)/\nu $,
 with  $m\,$ and $ m'\, $ being positive integers.
Denote for short $\fz = X^-_0$ and $\fy = X^+_{-1}$. The vector
$$  \cP_{(m,m';\nu )}V_{\{ J\,; \nu\}}
= \fz^{m+(m'-1)/\nu}\, \fy^{m+(m'-2)/\nu}\, \dots \,
  \fz^{m-(m'-3)/\nu}\, \fy^{m-(m'-2)/\nu}\, \fz^{m-(m'-1)/\nu}\,
V_{\{ J\,; \nu\}}\,, \eqno (31) $$
is a singular vector in the module generated on $V_{\{ J\,; \nu\}}\,$ [6]. It
can be cast in a canonical way into an integral powers
polynomial of $X^-_0$ and $X^a_{-n}$.
Furthermore  all the vectors
 obtained by
 dropping from the left the first one, or  the first two, etc.,
 of the  factors  in the r.h.s. of (31) , i.e.,
  $$   \cP_{(m,m';\nu)}V_{\{ J\,; \nu \}}  = V_{\{ J-m\,; \nu\}}=
  \fz^{m+(m'-1)/\nu}\, V_{\{- J-1+m\,; \nu\}}$$
$$ =
   \fz^{m+(m'-1)/\nu}\, \fy^{m+(m'-2)/\nu}\, V_{\{ J- m+1/\nu \,; \nu\}}
\,=\, ...   \, \,, $$
etc., formally have the properties of
 singular vectors.
 Hence they are annihilated by $Q$,  by the
positive modes of $\,T(z)\,, $  and furthermore they are  kept
invariant by the gauge transformation $\cA$.
We now apply $\cA$  from the left on both sides of (31)   using (29) and
identifying  subsequently  $V_0$ with the  subfactors. After
the first step we have (simplifying for short the notation
$\, \cO_{m} \equiv \cO_{(m;\nu)}$)

 $$\cP_{(m,m';\nu)}\,  V_{\{ J\,; \nu\}} = \cA\, \cP_{(m,m';\nu)}\,
 V_{\{ J\,; \nu\}} =
\cO_{m+(m'-1)/\nu} \, \fy^{m+(m'-2)/\nu}\,
  V_{\{ J- m+1/\nu \,; \nu\}} \,. \eqno(32)$$
Next  $\fy= X^+_{-1}\, $ can be rewritten as
  $\fy = 1+  \{b_0, Q \} \,$. Using that  $\,Q^2=0\,,$ any power
of $\fy \,$  can be written as
$\, (\fy )^n =  1+ Q \, A \, + \, A\, Q \,$ for some $A$.
Representing the power of $\fy \,$ in (32) in this way, the $Q\,$
 on the right will act on $V_{\{ J- m+1/\nu \,; \nu\}} \,$
  giving zero, while the $Q\,$ on the left passes through
$\cO_{m+(m'-1)/\nu} \,, $ producing $Q$ -exact terms.
 Thus we get for (32)

$$
=\cO_{m+(m'-1)/\nu}\,   \fz^{m+(m'-3)/\nu}\,
  \fy^{m+(m'-4)/\nu}\, \dots \,
\fz^{m-(m'-1)/\nu}\, V_{\{ J\,;\nu \}}\, + \, Q ...  $$
$$= \cO_{m+(m'-1)/\nu}\, \cA\,
  \fz^{m+(m'-3)/\nu}\,
 \,V_{\{- J-1+m- 1/\nu \,; \nu\}} + Q ... \, .$$

We repeat this step
 until $V_{\{ J\,; \nu\}}$ is reached,  turning
each power of $\fz$ in the MFF expression
  into the $\lbt$ - depending factors $\cO$, while getting rid of
the powers of $\fy$ at the price of $Q$ exact terms, i.e.,

$$\cP_{(m,m';\nu)} V_{\{ J\,; \nu\}} + Q ..... \, =
 \cO_{m+(m'-1)/\nu}\,\cO_{m+(m'-3)/\nu}
 \dots \cO_{m-(m'-1)/\nu}\,  V_{\{ J\,; \nu\}}  \equiv
 \cO_{(m,m'; \nu )}\,|h_{\{J\,; \nu \}}\rangle \,.\eqno(33)$$

We have   identified  the vector $V_{\{ J\,; \nu\}}$ with a  Vir
Verma module highest weight state $|h_{\{J\,; \nu \}}\rangle$.
The composition $\cO_{(m,m'; \nu )}
\,|h_{J\,; \nu}\rangle\,$ in  (33)
 as well as  any of its subfactors
are annihilated by positive grade Vir generators $\lbt_n\,, \ n\ge
1$, while $\lbt_0$ reproduces the correct eigenvalues.

Thus we have proved the statement

\medskip\noindent{\bf
 The general Vir singular
vectors $\cO_{(m,m'; \nu )}\,|h_{\{J\,; \nu \}}\rangle\,$
 written in the form of [9], are BRST equivalent to the MFF $\hs$
vectors $\cP_{(m,m';\nu)} V_{\{ J\,; \nu\}}$.}
\medskip

Note that the $\hs$ singular vectors of the type  $\fy^m V_{\{ J\,; \nu\}}\,,$
  $2J+1 = -m +1/\nu$,   are
BRST trivial, i.e.,   equivalent to $V_{\{ J\,;
 k\}}\,,$ in aggreement with the structure of  the corresponding
Vir modules.

The Vir singular  vector $\cO_{(m,m'; \nu )}\,|h_{\{J\,; \nu \}}\rangle\,$
 can be also recovered from  a singular vector of another $\hs$ module
labelled by the weight $ 2{ J'}+1 = 2/\nu - 2J -1 = -m + (m'+1) /\nu\,$. It is
given [6] by an expression like (31)  but
everywhere $\fz$  is replaced by $\fy$ and vice
versa, while the powers are kept the same as in (31)  with only
$m'$ changed to $m'+1$  and the h.w. state
 $V_{\{ J\,; \nu\}} $ replaced by  $V_{\{1/\nu -  J - 1 ;\nu\}}$.
On the other hand the invariance of $\,\{\, h_{\{J\,; \nu \}}\,, c_{\nu}\,
\} \,$
under the change
$\, \{ J\,, \nu \} \, \rightarrow \{-J \nu \,, 1/\nu \} \, $
of parameters  implies that
one and the same Vir singular vector admits different representations
 inherited from its $\,\hs\,$ counterparts.
 Finally when the $\,\hs\,$ vectors are true
 compositions (of factors with  positive integral powers, see [6]) they
 lead to Vir vectors expressible as a product of
 true BS-A factors.
\medskip
We conclude with the remark that the approch developed here can
be further exploited to derive more explicit
 formulae for the general Virasoro singular vectors,
starting directly from the  integral powers versions of the
MFF vectors.
\medskip

\bigskip\noindent{\bf Acknowledgements}
\medskip
We  acknowledge the financial support and  the hospitality  of INFN,
Sezione di Trieste and SISSA, Trieste.
\bigskip
\noindent{\bf References}
\medskip

\item{[1]}
     M. Bauer, Ph. Di Francesco, C. Itzykson and J.-B. Zuber,
    {\it Phys. Lett.} {\bf B260} (1991) 323;
    {\it Nucl. Phys.} {\bf B362} (1991) 515.

\item{[2]}
   L. Benoit and Y. Saint-Aubin, {\it Phys. Lett.} {\bf B215}
   (1988) 517.

\item{[3]}
 V. Drinfeld and V. Sokolov, {\it J. Sov. Math.} {\bf 30} (1984) 1975.

\item{[4]}
   P. Furlan, R. Paunov, A.Ch. Ganchev and V.B. Petkova,
    {\it Nucl. Phys.} {\bf B394} (1993) 665.

\item{[5]}
   A.Ch. Ganchev and V.B. Petkova,
{\it Phys. Lett.} {\bf B293}   (1992) 56.

\item{[6]}
   F.G. Malikov, B.L. Feigin and D.B. Fuks, {\it Funct. Anal.
   Prilozhen.} {\bf 20}, no.  2 (1987) 25.

\item{[7]}
   M. Bershadsky and H. Ooguri, {\it Comm. Math. Phys.} {\bf
  126} (1989) 49.

\item{[8]}
 M. Bauer and N. Sochen,
{\it Comm. Math. Phys.} {\bf 152 } (1993) 127.

\item{[9]}
    A. Kent, Singular vectors of the Virasoro algebra,
 {\it Phys. Lett.} {\bf B273} (1991) 56.

\item{[10]}
   P. Furlan,  A.Ch. Ganchev and V.B. Petkova,
   $W_3$ singular vectors via quantum DS reduction, preprint
   Trieste,  in preparation.

\item{[11]}
       F.A. Bais, P. Bouwknegt, M. Surridge and K. Schoutens,
    {\it Nucl. Phys.} {\bf B304} (1988) 348.

\item{[12]}
   B. Feigin and E. Frenkel, {\it Phys. Lett.} {\bf B246} (1990) 75.

\item{[13]}
  J. Boer and T. Tjin, The relation between quantum $W$ algebras
and Lie algebras, preprint THU - 93/05, ITFA - 02 - 93.

\bye